# Sentiment Analysis of Review Datasets using Naïve Bayes' and K-NN Classifier

Lopamudra Dey
Department of Computer Science & Engineering
Heritage Institute of Technology
Kolkata, India
Email: lopamudra.dey@heritageit.edu

Sanjay Chakraborty
Department of Computer Science & Engineering
Institute of Engineering & Management
Kolkata, India
Email:sanjay.chakraborty@iemcal.com

Anuraag Biswas
Computer Science & Engineering
Heritage Institute of Technology
Kolkata, India
Email:anuraagbiswas111@gmail.com

Beepa Bose
Computer Science & Engineering
Heritage Institute of Technology
Kolkata, India
Email:beepabose@gmail.com

Sweta Tiwari
Computer Science & Engineering
Heritage Institute of Technology
Kolkata, India
Email:sweta.tiwari604@gmail.com

*Abstract*—The advent of Web 2.0 has led to an increase in the amount of sentimental content available in the Web. Such content is often found in social media web sites in the form of movie or product reviews, user comments, testimonials, messages in discussion forums etc. Timely discovery of the sentimental or opinionated web content has a number of advantages, the most important of all being monetization. Understanding of the sentiments of human masses towards different entities and products enables better services for contextual advertisements, recommendation systems and analysis of market trends. The focus of our project is sentiment focussed web crawling framework to facilitate the quick discovery of sentimental contents of movie reviews and hotel reviews and analysis of the same. We use statistical methods to capture elements of subjective style and the sentence polarity. The paper elaborately discusses two supervised machine learning algorithms: K-Nearest Neighbour(K-NN) and Naïve Bayes' and compares their overall accuracy, precisions as well as recall values. It was seen that in case of movie reviews Naïve Bayes' gave far better results than K-NN but for hotel reviews these algorithms gave lesser, almost same accuracies.

## I. INTRODUCTION

Data mining is a process of mined valuable data from a large set of data. Several analysis tools of data mining (like, clustering, classification, regression etc,) can be used for sentiment analysis task [13][14]. Sentiment mining is one of the important aspects of data mining where important data can be mined based on the positive or negative senses of the collected data. Sentiment Analysis also known as Opinion Mining refers to the use of natural language processing, text analysis and computational linguistic to identify and extract subjective information in source materials.
Here the source materials refer to opinions / reviews /comments given in various social networking sites [1].The Sentiment found within comments, feedback or critiques provide useful indicators for many different purposes and can be categorized by polarity [2].By polarity we tend to find out if a review is overall a positive one or a negative one. For example:

  1) Positive Sentiment in subjective sentence: "I loved the movie Mary Kom"—This sentence is expressed positive sentiment about the movie Mary Kom and we can decide that from the sentiment threshold value of word "loved". So, threshold value of word "loved" has positive numerical threshold value.

  2) Negative sentiment in subjective sentences: "Phata poster nikla hero is a flop movie" defined sentence is expressed negative sentiment about the movie named "Phata poster nikla hero" and we can decide that from the sentiment threshold value of word "flop". So, threshold value of word "flop" has negative numerical threshold value. Sentiment Analysis is of three different types: Document level, Sentence level and Entity level. However we are studying phrase level sentiment analysis. The traditional text mining concentrates on analysis of facts whereas opinion mining deals with the attitudes [3]. The main fields of research are sentiment classification, feature based sentiment classification and opinion summarizing. Now, the use of sentiment analysis in a commercial environment is growing. This is evident in the increasing number of

brand tracking and marketing companies offering this service. Some services include:

to prevent viral effects.
- Assessing market buzz, competitor activity and customer trends, fads and fashion.
- Measuring public response to an activity or company related issue [4].

In this paper for Sentiment Analysis we are using two Supervised Machine Learning algorithms : Naïve Bayes' and K-Nearest Neighbour to calculate the accuracies, precisions (of positive and negative corpuses) and recall values (of positive and negative corpuses). The difficulties in Sentiment Analysis are an opinion word which is treated as positive side may be considered as negative in another situation. Also the degree of positivity or negativity also has a great impact on the opinions. For example "good" and "very good" cannot be treated same.[2] Although the traditional text processing says that a small change in two pieces of text does not change the meaning of the sentences. However the latest text mining gives room for advanced analysis measuring the intensity of the word. Here is the point where we can scale the accuracy and efficiency of different algorithms [4].

The rest of the paper is organized as follows: Section 2 deals with the related work of our study, Section 3 presents our proposed work (Data sets and data sources used in our study along with the models and methodology used), Section 4 presents all our experimental results, Section 5 presents the conclusion drawn from our survey.

## II. RELATED WORK

Several techniques were used for Sentiment Analysis. Few Related work are as follow:
(a)Mori Rimon[3] used the keyword based approach to classify sentiment. He worked on identifying keywords basically adjectives which indicates the sentiment. Such indicators can be prepared manually or derived from Wordnet.
(b)Alec co [4] used different machine learning algorithms such as Naïve Bayes', Support vector machine and maximum entropy.
(c)Janice M. Weibe [5] performed document and sentence level classification. He fetched review data from different product destinations such as automobiles, banks, movies and travel. He classified the words into positive and negative categories. He then calculated the overall positive or negative score for the text. If the number of positive words is more than negative then the document is considered positive otherwise negative.
(d) Jalaj S. Modha , Gayatri S. Pandi and Sandip J. Modha [6] worked on techniques of handling both subjective as well as objective unstructured data.
(e) Theresa Wilson, Janyce Wiebe and Paul Hoffman [7] worked on a new approach on sentiment analysis by first determining whether an expression is neutral or

- Tracking users and non-users opinions and ratings on products and services.
- Monitoring issues confronting the company so as polar and then disambiguates the polarity of the polar expression. With this approach the system is able to automatically identify the contextual polarity for a large subset of sentiment expressions, hence achieving results which are better than baseline.

## II. PROPOSED WORK

*A) Data source and Data Set*
To conduct the research, two datasets are considered here - Movie Reviews & Hotel Reviews.
- All the movie reviews have been scanned from www.imdb.com.
- All the hotel reviews have been downloaded from OpinRank Review Dataset (http://archive.ics.uci.edu/ml/datasets/OpinRank+Review+Dataset)

The data set has been prepared by taking 5000 positive and 5000 negative reviews from each of the mentioned sites.

*B) Methodology*

The main goal of the research is to analyse the data from the surveys and to decide whether it is suitable to be analysed with the use of the discussed data mining methods. A graphical description of the processes involve in sentiment analysis is detailed in Figure 1 below.

Fig 1. Sentiment Analysis Process

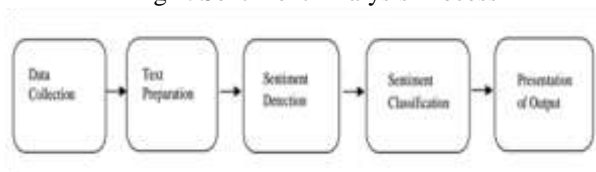

### 1) Naïve Bayes' Classifier

Bayesian network classifiers are a popular supervised classification paradigm. A well-known Bayesian network classifier is the Naïve Bayes' classifier is a probabilistic classifier based on the Bayes' theorem, considering Naïve (Strong) independence assumption. It was introduced under a different name into the text retrieval community and remains a popular(baseline) method for text categorizing, the problem of judging documents as belonging to one category or the other with word frequencies as the feature. An advantage of Naïve Bayes' is that it only requires a small amount of training data to estimate the parameters necessary for classification. Abstractly, Naïve Bayes' is a conditional probability model. Despite its simplicity and strong assumptions, the naïve Bayes' classifier has been proven to work satisfactorily in many domains. Bayesian classification provides practical learning algorithms and prior knowledge and observed data can be combined. In Naïve probabilities of categories given a text document by using the joint probabilities of words and categories. It is based on the assumption of word independence. The starting point is the Bayes' theorem for conditional probability, stating that, for a given data point x and class C:

$$P(C/x) = P(x/C)/P(x) \qquad (1)$$

Furthermore, by making the assumption that for a data point $x = \{x_1, x_2, ... x_j\}$, the probability of each of its attributes occurring in a given class is independent, we can estimate the probability of x as follows:

$$P(C/x) = P(C) . \prod P(x_i/C) \qquad (2)$$

## Algorithm

*Input*: a document *d*
A fixed set of classes $C = \{c_1, c_2, ..., c_j\}$
*Output*: a predicted class $c \in C$

**Steps:**
1. *Pre-processing:*
i. About 10,000 reviews were crawled from www.imdb.com / OpinRank Review Dataset
ii. Positive reviews and negative reviews were kept in two files pos.txt and neg.txt
iii. 2 empty lists were taken, one for positive and one for negative reviews.
iv. Sentences of the positive and negative reviews were broken and 'pos' and 'neg' were appended to each accordingly and were stored in the 2 empty lists created.
v. ¾ of these sentences were kept in the dictionary for training while the ¼ were kept for testing.
2. Using chi squared test (explained later) we calculated the score of each of the remaining words and instead of using all of those words we only used the best 10,000.
3. The classifier was trained using the dataset just prepared.
4. Labelled sentences were kept correctly in reference sets and the predicatively labelled version in test sets.
5. Metrics were calculated accordingly.

Bayes' technique, the basic idea to find the

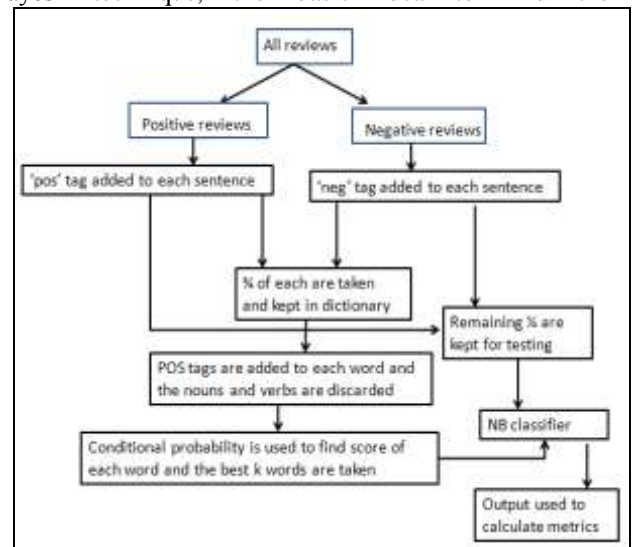

Fig 2. Naïve Bayes' flowchart

A small example using Naïve Bayes' is given below,

| Set | Document | Review Sentence | Class |
|---|---|---|---|
| Training Set | 1 | I liked the movie | pos |
| | 2 | It's a good movie. Nice story. | pos |
| | 3 | Hero's acting is bad but heroine looks good. Overall nice movie. | pos |
| | 4 | Nice songs. But sadly boring ending. | neg |
| Test Set | | I like the direction. But boring locations. Overall good movie | pos |

*2) k-Nearest Neighbour Classifier*

K-NN is a type of instance-based learning, or lazy learning where the function is only approximated locally and all computation is deferred until classification. It is non parametric method used for classification or regression. In case of classification the output is class membership (the most prevalent cluster may be returned), the object is classified by a majority vote of its neighbours, with the object being assigned to the class most common among its *k nearest neighbours*. This rule simply retains the entire training set during learning and assigns to each query a class represented by the majority label of its k-nearest neighbours in the training set.

The *Nearest Neighbour* rule (NN) is the simplest form of K-NN when K = 1. Given an unknown sample and a training set, all the distances between the unknown sample and all the samples in the training set can be computed. The distance with the smallest value corresponds to the sample in the training set closest to the unknown sample. Therefore, the unknown sample may be classified based on the classification of this nearest neighbour. The K-NN is an easy algorithm to understand and implement, and a powerful tool we have at our disposal for sentiment analysis. K-NN is powerful because it does not assume anything about the data, other than a distance measure can be calculated consistently between two instances. As such, it is called non-parametric or non-linear as it does not assume a functional form. The flowchart of k-nn classifier is given in Fig.3.

**Algorithm:**

*1. Pre-processing:*
i). About 10,000 reviews were crawled from **www.imdb.com/OpinRank Review Dataset**
ii. Positive reviews and negative reviews were kept in two files pos.txt and neg.txt
iii. 2 empty lists were taken, one for positive and one for negative reviews.
iv. Sentences of the positive and negative reviews were broken and 'pos' and 'neg' were appended to each accordingly and were stored in the 2 empty lists created.
v. ¾ of these sentences were kept in the dictionary for training while the ¼ were kept for testing.

*2. Training:*
i. Using chi squared test we calculated the score of each of the words occurring in the training dataset.
ii. An empty list is created, the dictionary in which the words from training dataset are stored followed by each of their scores thus calculated.
ii. for each test review
iii. for each word
iv. If it exists in the word score list, add its score to review score
v. Else find the word in word score list with minimum jaccard index to the unknown word and add its score to the review score.
vi. End for at step 3
vii. End for at step 4
viii. Find metrics accordingly.

*Chi squared test:*

1. Initialize an empty frequency distribution.
2. Initialize an empty conditional frequency distribution (based on words being positive and negative).
3. We fill out the frequency distributions, incrementing the counter of each word within the appropriate distribution.
4. We find the highest-information features is the count of words in positive reviews, words in negative reviews, and total words.
5. We use a chi-squared test (also from NLTK) to score the words. We find each word's positive information score and negative information score, add them up, and fill up a dictionary correlating the words and scores, which we then return out of the function.

IV. EXPERIMENTAL RESULTS

Accuracy, Precision and recall are method used for evaluating the performance of opinion mining. Here accuracy is the overall accuracy of certain sentiment models. Recall (Pos) and Precision (Pos) are the ratio and precision ratio for true positive reviews. Recall (Neg) and Precision (Neg) are the ratio and precision ratio for true negative reviews. In an ideal scenario, all the experimental results are measured according to the Table 1.and accuracy, Precision and recall as explained below [9].

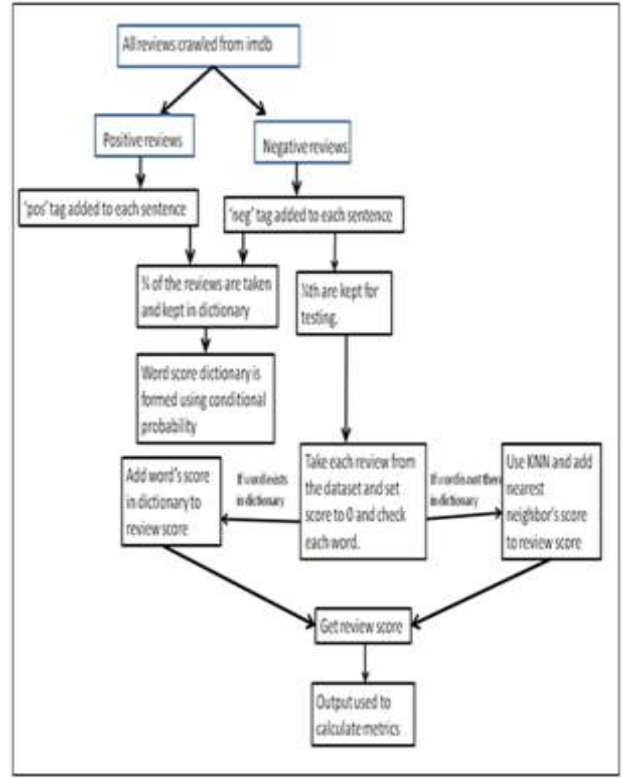

Fig 3. K-nn Classifier flowchart

$$Accuracy = \frac{a+d}{a+b+c+d}$$

$$Recall(Pos) = \frac{a}{a+c} \qquad Recall(Neg) = \frac{d}{b+d}$$

$$Precision(Pos) = \frac{a}{a+b} \qquad Precision(Neg) = \frac{d}{c+d}$$

Table 1. A confusion Table

|  | True positive reviews | True negative reviews |
|---|---|---|
| Predict positive reviews | a | b |
| Predict negative reviews | c | d |

The overall accuracies of the three algorithms in 10 rounds of experiments are indicated in Table 2 and Fig.4,

Table 2. Accuracy comparison on Test Datasets.

| No. Of experiments | Number of reviews in the training dataset | Accuracy | | | |
|---|---|---|---|---|---|
| | | Naïve Bayes' (movie reviews) | K-NN (movie reviews) | Naïve Bayes' (hotel reviews) | K-NN (hotel reviews) |
| 1. | 100 | 56.78 | 47.64 | 43.11 | 45.35 |
| 2. | 200 | 64.29 | 55.07 | 41.26 | 40.97 |
| 3. | 500 | 70.06 | 58.44 | 42.56 | 41.42 |
| 4. | 1000 | 73.81 | 61.48 | 44.64 | 41.18 |
| 5. | 1500 | 77.23 | 64.21 | 48.21 | 42.01 |
| 6. | 2000 | 79.14 | 66.02 | 51.28 | 46.57 |
| 7. | 2500 | 79.82 | 67.89 | 52.03 | 47.04 |
| 8. | 3000 | 80.27 | 68.58 | 52.64 | 47.03 |
| 9. | 4000 | 82.11 | 69.03 | 53.92 | 49.75 |
| 10. | 4500 | 82.43 | 69.81 | 55.09 | 52.14 |

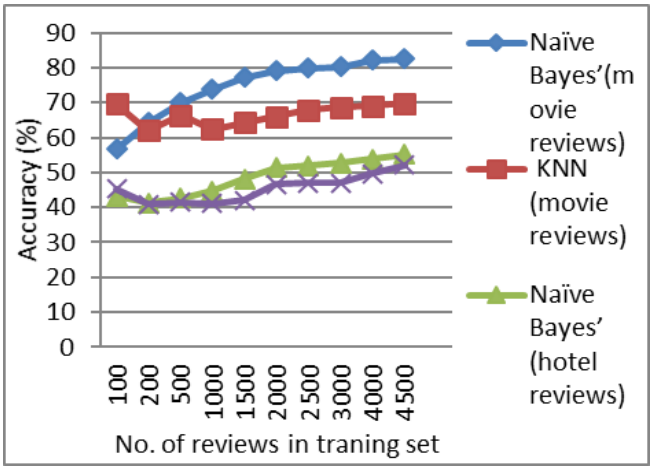

Fig. 4. Diagrammatic presentation of accuracies in the experiments

Table 3. Result of accuracies with maximum number of reviews:

| Total number of reviews | Classifier used | Review dataset used | Correct Sample | Incorrect Sample |
|---|---|---|---|---|
| 1500 | Naïve Bayes' | Movies | 1237 | 263 |
| | | Hotel | 827 | 673 |
| | K-NN | Movies | 1047 | 453 |
| | | Hotel | 782 | 718 |

Table 4. Precision comparison for Positive Corpus on Test Datasets

| No. Of experiments | Number of reviews in the training dataset | Precision for positive corpus: | | | |
|---|---|---|---|---|---|
| | | Naïve Bayes' (movie reviews) | K-NN (movie reviews) | Naïve Bayes' (hotel reviews) | K-NN (hotel reviews) |
| 1. | 100 | 59.04 | 41.35 | 42.11 | 44.51 |
| 2. | 200 | 64.96 | 50.97 | 40.26 | 40.86 |
| 3. | 500 | 69.56 | 54.42 | 41.56 | 40.41 |
| 4. | 1000 | 73.64 | 58.18 | 43.64 | 42.21 |
| 5. | 1500 | 77.21 | 62.01 | 47.21 | 42.12 |
| 6. | 2000 | 80.28 | 65.57 | 50.28 | 45.36 |
| 7. | 2500 | 81.03 | 66.04 | 51.03 | 46.14 |
| 8. | 3000 | 81.64 | 67.03 | 51.64 | 47.13 |
| 9. | 4000 | 82.92 | 67.75 | 52.92 | 47.57 |
| 10. | 4500 | 84.09 | 68.14 | 54.09 | 48.21 |

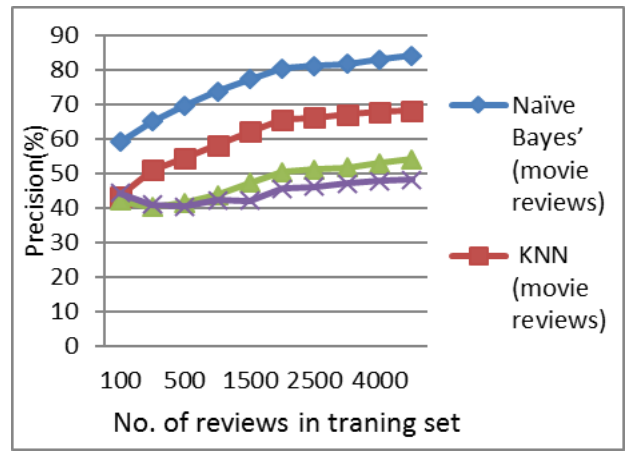

Fig 5. Diagrammatic presentation of positive precisions in the experiments

Table 5. Precision comparison for Negative Corpus on Test Datasets

| No. Of experiments | Number of reviews in the training dataset | Precision for negative corpus: | | | |
|---|---|---|---|---|---|
| | | Naïve Bayes' (movie reviews) | K-NN (movie reviews) | Naïve Bayes' (hotel reviews) | K-NN (hotel reviews) |
| 1. | 100 | 55.43 | 38.12 | 48.39 | 46.21 |
| 2. | 200 | 63.67 | 49.56 | 42.61 | 41.63 |
| 3. | 500 | 70.59 | 57.25 | 50.62 | 47.32 |
| 4. | 1000 | 73.99 | 62.12 | 53.81 | 52.15 |
| 5. | 1500 | 77.25 | 64.48 | 57.31 | 54.43 |
| 6. | 2000 | 78.09 | 65.73 | 58.11 | 55.69 |
| 7. | 2500 | 78.70 | 66.23 | 58.4 | 56.32 |
| 8. | 3000 | 79.00 | 66.47 | 59.91 | 56.51 |
| 9. | 4000 | 81.33 | 66.62 | 61.29 | 56.66 |
| 10. | 4500 | 81.01 | 66.73 | 61.11 | 56.77 |

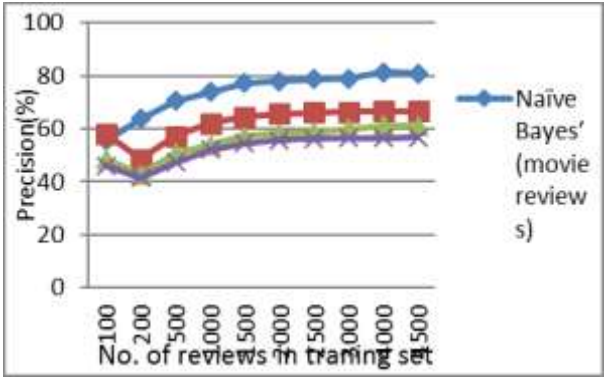

Fig 6. Diagrammatic presentation negative precisions in the experiments

Table 6. Recall Comparison for positive corpus on test datasets

| No. Of experiments | Number of reviews in the training dataset | Recall for positive corpus: | | | |
|---|---|---|---|---|---|
| | | Naïve Bayes' (movie reviews) | K-NN (movie reviews) | Naïve Bayes' (hotel reviews) | K-NN (hotel reviews) |
| 1. | 100 | 44.33 | 31.12 | 32.24 | 30.35 |
| 2. | 200 | 62.04 | 45.37 | 43.54 | 42.41 |
| 3. | 500 | 71.34 | 52.24 | 41.79 | 41.86 |
| 4. | 1000 | 74.19 | 56.31 | 47.44 | 42.21 |
| 5. | 1500 | 77.26 | 58.24 | 49.19 | 44.72 |
| 6. | 2000 | 77.26 | 60.02 | 50.02 | 45.03 |
| 7. | 2500 | 77.89 | 61.12 | 51.77 | 46.01 |
| 8. | 3000 | 78.09 | 61.53 | 51.44 | 46.52 |
| 9. | 4000 | 80.87 | 61.72 | 51.34 | 46.25 |
| 10. | 4500 | 80.12 | 61.81 | 51.84 | 46.31 |

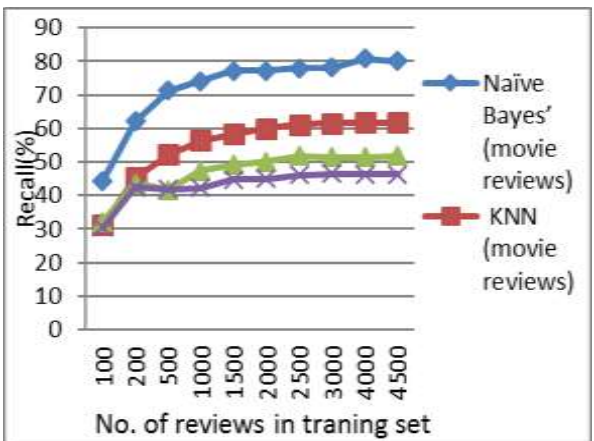

Fig 7. Diagrammatic presentation of recall for positive corpus in the experiments

Table 7. Recall Comparison for negative corpus on test datasets

| No. Of experiments | Number of reviews in the training dataset | Recall for negative corpus in test dataset | | | |
|---|---|---|---|---|---|
| | | Naïve Bayes' (movie reviews) | K-NN (movie reviews) | Naïve Bayes' (hotel reviews) | K-NN (hotel reviews) |
| 1. | 100 | 69.24 | 39.25 | 62.33 | 60.35 |
| 2. | 200 | 66.54 | 55.12 | 53.51 | 52.41 |
| 3. | 500 | 68.79 | 53.86 | 51.81 | 51.89 |
| 4. | 1000 | 73.44 | 60.21 | 57.52 | 52.19 |
| 5. | 1500 | 77.19 | 63.72 | 59.24 | 54.77 |
| 6. | 2000 | 81.02 | 65.03 | 60.11 | 5513 |
| 7. | 2500 | 81.77 | 66.01 | 61.83 | 56.11 |
| 8. | 3000 | 82.44 | 66.52 | 61.49 | 56.32 |
| 9. | 4000 | 83.34 | 66.25 | 61.37 | 56.35 |
| 10. | 4500 | 84.84 | 66.31 | 61.88 | 56.41 |

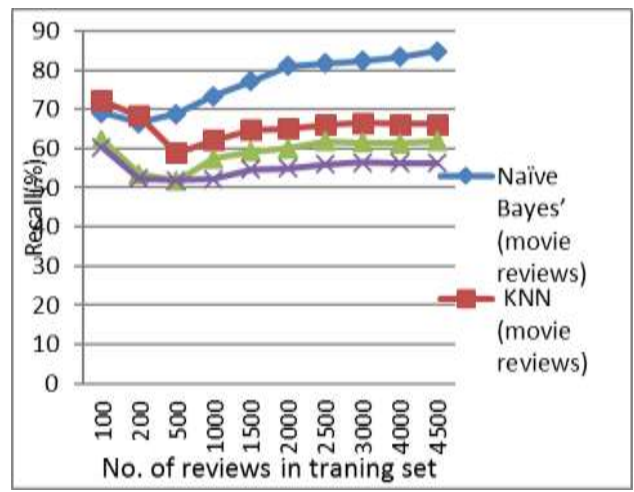

Fig 8. Diagrammatic presentation of recall for negative corpus in the experiments

## V. CONCLUSION

The aim of study is to evaluate the performance for sentiment classification in terms of accuracy, precision and recall. In this paper, we compared two supervised machine learning algorithms of Naïve Bayes' and K-NN for sentiment classification of the movie reviews and hotel reviews. The experimental results show that the classifiers yielded better results for the movie reviews with the Naïve Bayes' approach giving above 80% accuracies and outperforming than the k-NN approach. However for the hotel reviews, the accuracies are much lower and both the classifiers yielded similar results. Thus we can say Naïve Bayes' classifier can be used successfully to analyse movie reviews.

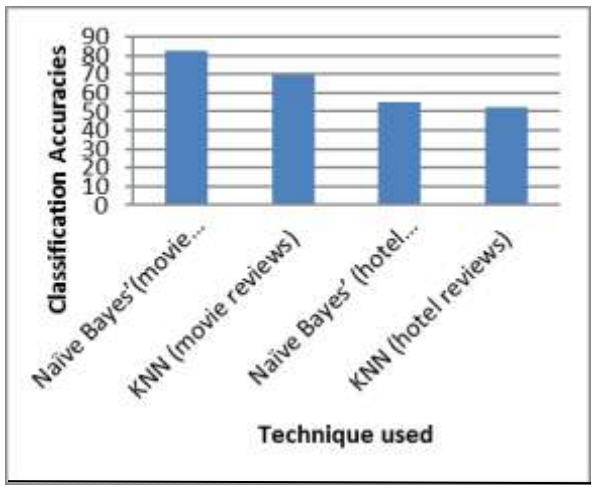

Fig 9. Accuracies of the classifiers with the 2 datasets

VI. FUTURE WORK

For further work we would like to compare try and come up with an efficient sentiment analyser like random forest, Support vector Machine etc. And also try to implement a new algorithm utilizing the benefits of the both algorithms so that it can be used effectively in data forecasting.